\documentstyle[12pt,aps]{revtex}
\def\nc{\newcommand}
\nc{\bA}{\mbox{\boldmath $A$\unboldmath}}
\nc{\bn}{\mbox{\boldmath $n$\unboldmath}}
\nc{\bl}{\mbox{\boldmath $l$\unboldmath}}
\nc{\bm}{\mbox{\boldmath $m$\unboldmath}}

\def\p{\partial}
\nc{\pr}{\frac{\p}{\p r}}
\nc{\pv}{\frac{\p}{\p v}}
\nc{\pta}{\frac{\p}{\p \theta}}
\nc{\pvi}{\frac{\p}{\p \varphi}}

\nc{\pdr}{\frac{\p^2}{\p r^2}}
\nc{\pdv}{\frac{\p^2}{\p v^2}}
\nc{\pdta}{\frac{\p^2}{\p \theta^2}}
\nc{\pdvi}{\frac{\p^2}{\p \varphi^2}}
\nc{\pdva}{\frac{\p^2}{\p v \p \varphi}}
\nc{\pdvr}{\frac{\p^2}{\p v \p r}}
\nc{\pdri}{\frac{\p^2}{\p r \p \varphi}}

\nc{\spr}{\frac{\p}{\p r_*}}
\nc{\spv}{\frac{\p}{\p v_*}}
\nc{\spta}{\frac{\p}{\p \theta_*}}

\nc{\spdr}{\frac{\p^2}{\p r_*^2}}
\nc{\spdvr}{\frac{\p^2}{\p r_* \p v_*}}
\nc{\spdra}{\frac{\p^2}{\p r_* \p \theta_*}}
\nc{\spdri}{\frac{\p^2}{\p r_* \p \varphi}}

\nc{\sta}{\sin\theta}
\nc{\cta}{\cos\theta}
\nc{\sda}{\sin^2\theta}
\nc{\cda}{\cos^2\theta}
\nc{\coa}{\cot\theta}
\nc{\sqd}{\sqrt{2}}
\nc{\cD}{\cal D}
\nc{\cL}{\cal L}
\nc{\cLd}{{\cal L}^{\dagger}}
\nc{\drH}{\dot{r}_H}
\nc{\ddrH}{\ddot{r}_H}
\nc{\prH}{r_H^{\prime}}
\nc{\ka}{\kappa}
\nc{\tkr}{2\kappa(r-r_H)}
\nc{\pprH}{r_H^{\prime\prime}}

\begin{document}
\baselineskip 16pt

\title{Hawking Radiation of a Non-stationary
Kerr-Newman Black Hole: Spin-Rotation Coupling Effect}
\author{S. Q. Wu\thanks{E-mail: sqwu@iopp.ccnu.edu.cn}
and X. Cai\thanks{E-mail: xcai@ccnu.edu.cn}}
\address{Institute of Particle Physics, Hua-Zhong
Normal University, Wuhan 430079, P.R. China}
\maketitle

\bigskip
\begin{quote}
Hawking evaporation of Klein-Gordon and Dirac particles in a
non-stationary Kerr-Newman space-time is investigated by using
a method of generalized tortoise coordinate transformation.
The location and the temperature of the event horizon of a
non-stationary Kerr-Newman black hole are derived. It is shown
that the temperature and the shape of the event horizon depend not
only on the time but also on the angle. However, the Fermionic
spectrum of Dirac particles displays a new spin-rotation coupling
effect which is absent from that of Bosonic distribution of
scalar particles. The character of this effect is its obvious
dependence on different helicity states of particles spin-$1/2$.

\end{quote}

\section{Introduction}

The fourth quarter of last century has witnessed various remarkable
progress on several researches on black hole physics since Hawking's
remarkable discovery \cite{Hawk}. One of these aspects is to reveal
the thermal properties of many kinds of black holes by miscellaneous
methods (\cite{FN,Zhao} and references therein). Much efforts have been
devoted to the thermal radiation of scalar and Dirac fields in some
static or stationary black holes \cite{FN,Zhao,HR,AMKP,WC1}. In the
case of a non-stationary axisymmetric black hole, many works have
been done on the radiation of scalar particles \cite{Zhao,KGNK}.
Though Hawking effect of Dirac particles has been investigated in
the non-static case \cite{Zhaoel}, it is difficult to deal with
the Hawking evaporation of Dirac particles in a non-stationary
axisymmetric black hole. The difficulty lies in the non-separability
of the radial and angular variables for Chandrasekhar-Dirac equations
\cite{CP} in the non-stationary axisymmetry space-time. Recently this
dilemma has been attacked by us \cite{WC2} through considering
simultaneously the asymptotic behaviors of the first-order and
second-order forms of Dirac equation near the event horizon. A
new term representing the interaction between the spin of Dirac
particles and the angular momentum of evaporating Kerr black holes
was observed in the thermal radiation spectrum of Dirac particles.
The character of this spin-rotation coupling effect is its obvious
dependence on different helicity states of particles with spin-$1/2$.
This effect disappears \cite{WC3} when the space-time degenerates
to a spherically symmetric black hole of Vaidya-type. It should be
noted that this term displayed in the Fermi-Dirac spectrum is
absent in the Bose-Einstein distribution of Klein-Gordon particles.

In this paper, we extend the method developed in Ref. \cite{WC2}
to tackle with the thermal radiation of Klein-Gordon and Dirac
particles in a non-stationary Kerr-Newman space-time. This analysis
is of theoretical interest under current consideration. It is shown
that the location and the temperature of the event horizon depend on
the time and the angle. The Fermionic spectrum of Dirac particles
displays a spin-rotation coupling effect due to the interaction
between the particles with spin-$1/2$ and the black holes with
rotation.

The paper is outlined as follows: In Sec. 2, the location of the
event horizon of a non-stationary Kerr-Newman black hole is derived
by using the method of generalized tortoise coordinate transformation.
Then Klein-Gordon equation of scalar particles and Dirac equation of
spinor fields are manipulated in Sec. 3 and Sec. 4, respectively. In
Sec. 5, both equations for massive particles are recast into a standard
wave equation near the event horizon, and the \lq\lq surface gravity"
of the event horizon is obtained. Sec. 6 is devoted to derive the
thermal radiation spectra of scalar and spinor particles. In Sec. 7,
we present some discussions about the spin-rotation coupling effect.

\section{Generalized Tortoise
Coordinate Transformation Method}

The metric of a non-stationary Kerr-Newman black hole \cite{GHJW,CKC}
can be written in the advanced Eddington-Finkelstein system as
\begin{eqnarray}
&ds^2& = \frac{\Delta-a^2\sda}{\Sigma}dv^2 +2\frac{r^2+a^2
-\Delta}{\Sigma}a\sda dvd\varphi -2dvdr  \nonumber\\
&&+2a\sda drd\varphi -\Sigma d\theta^2 -\frac{(r^2+a^2)^2
-\Delta a^2\sda}{\Sigma}\sda d\varphi^2 \, , \label{KNbh}
\end{eqnarray}
where $\Delta=r^2-2M(v)r+Q^2(v)+a^2$, $\Sigma=r^2+a^2\cda=\rho^*\rho$,
$\rho^*=r+ia\cta$, $\rho=r-ia\cta$, and $v$ is the standard advanced time.
Both the mass $M$ and the charge $Q$ of the hole depend on the time $v$,
but the specific angular momentum $a$ is a constant. The metric (\ref{KNbh})
and the electro-magnetic potential
\begin{equation}
{\cal A}=\frac{Q(v)r}{\Sigma}\Big(dv -a\sda d\varphi\Big) \, ,
\end{equation}
is shown by Xu \cite{XDY} to be an exact solution of the
Einstein-Maxwell equations.

The line element (\ref{KNbh}) of an evaporating Kerr-Newman black hole
is a natural non-stationary generalization of the stationary Kerr-Newman
solution, but it is of Petrov type-II, whereas the latter is of Petrov
type-D. The geometry of this space-time is characterized by three kinds
of surfaces of particular interest: the apparent horizons $r_{AH}^{\pm}
= M\pm (M^2-Q^2-a^2)^{1/2}$, the timelike limit surfaces $r_{TLS}^{\pm}
= M\pm (M^2-Q^2-a^2\cda)^{1/2}$, and the event horizons $r_{EH}^{\pm}
= r_H$. The event horizon is necessarily a null-surface $r = r(v,\theta)$
that satisfies the null-surface conditions $g^{ij}\p_i F\p_j F = 0$ and
$F(v,r,\theta) = 0$. An effective method to determine the location and
the temperature of the event horizon of a dynamic black hole is called
the generalized tortoise coordinate transformation (GTCT) which can give
simultaneously the exact values both of the location and of the
temperature of the event horizon of a non-stationary black hole.
Basically, this method is to reduce Klein-Gordon or Dirac equation
in a known black hole background to a standard wave equation near
the event horizon by generalizing the common tortoise-type coordinate
$r_* = r +(2\ka)^{-1}\ln(r -r_H)$ in a static or stationary space-time
\cite{DRS} (where $\kappa$ is the surface gravity of the studied event
horizon) to a similar form in a non-static or non-stationary space-time
and by allowing the location of the event horizon $r_H$ to be a function
of the advanced time $v = t +r_*$ and/or the angles $\theta,\varphi$.

As the space-time under consideration is symmetric about
$\varphi$-axis, one can introduce the following generalized
tortoise coordinate transformation (GTCT) \cite{WC2}
\begin{eqnarray}
&&r_*=r +\frac{1}{2\ka(v_0,\theta_0)}\ln[r-r_H(v,\theta)]
\, , \nonumber\\
&&v_*=v-v_0 \, , ~~~~~~~~~~\theta_*=\theta -\theta_0 \, ,
\label{trans}
\end{eqnarray}
where $r_H = r(v,\theta)$ is the location of event horizon, and $\ka$ is
an adjustable parameter. All parameters $\kappa$, $v_0$ and $\theta_0$
characterize the initial state of the hole and are constant under the
tortoise transformation.

Applying the GTCT of Eq. (\ref{trans}) to the null surface
equation $g^{ij}\p_i F\p_j F=0$ and then taking the $r \rightarrow
r_H(v_0,\theta_0)$, $v \rightarrow v_0$ and $\theta \rightarrow
\theta_0$ limits, we arrive at
\begin{equation}
\Big[\Delta_H -2(r_H^2 +a^2)\drH +a^2\sda_0 {\drH}^2
+{\prH}^2\Big]\Big(\spr F\Big)^2 = 0 \, ,
\end{equation}
in which the vanishing of the coefficient in the square bracket can
give the following equation to determine the location of the event
horizon of an evaporating Kerr-Newman black hole
\begin{equation}
\Delta_H -2(r_H^2 +a^2)\drH
+a^2\sda_0 {\drH}^2 +{\prH}^2 = 0 \, ,
\label{loeh}
\end{equation}
where we denote $\Delta_H = r_H^2 -2Mr_H -Q^2+a^2$. The
quantities $\drH=\p r_H/\p v$ and $\prH=\p r_H/\p \theta$ depict
the change of the event horizon in the advanced time and with the
angle, which reflect the presence of quantum ergosphere near the
event horizon. Eq. (\ref{loeh}) means that the location of the
event horizon is shown as
\begin{equation}
r_H=\frac{M}{1-2\drH}\pm \Big[\frac{M^2}{(1-2\drH)^2}
-\frac{Q^2 +a^2\sda_0{\drH}^2 +{\prH}^2}{1-2\drH}
-a^2\Big]^{1/2} \, . \label{loca}
\end{equation}
The plus (minus) sign corresponds to an outer (inner) event horizon.

\section{Klein-Gordon Equation}

In this section, we will consider the asymptotic behavior of minimally
electro-magnetic coupling Klein-Gordon equation near the event horizon.
The explicit form of wave equation describing the dynamic behavior of
scalar particles with mass $\mu_0$ and charge $q$
\begin{equation}
\frac{1}{\sqrt{-g}}(\p_k+iqA_k)\Big[\sqrt{-g}g^{kj}(\p_j+iqA_j)\Phi
\Big]+\mu_0^2\Phi = 0 \, ,
\end{equation}
in the above space-time (\ref{KNbh}) is
\begin{eqnarray}
&&\Big[\Delta\pdr +2(r^2+a^2)\pdvr +2a\pdri +2a\pdva +\pdta \nonumber\\
&&~\quad +\frac{1}{\sda}\pdvi +a^2\sda\pdv +\coa\pta +2r\pv \nonumber\\
&&~~\quad\quad +2(r-M+iqQr)\pr +iqQ-\mu_0^2\Sigma\Big]\Phi=0 \, .
\end{eqnarray}
Under the GTCT (\ref{trans}), it can be transformed into
\begin{eqnarray}
&&\Big[\frac{r_H(1 -2\drH) -M}{\ka} +2\Delta_H
-2\drH (r_H^2+a^2)\Big]\spdr \Phi -2\prH \spdra \Phi  \nonumber \\
&&~ +2a(1 -\drH) \spdri \Phi +2(r_H^2 +a^2 -\drH a^2\sda_0) \spdvr \Phi
\nonumber \\
&&~ -(-2r_H\drH -2iqQr_H +\prH\coa_0 +\pprH +\ddrH a^2\sda_0)\spr\Phi = 0 \, .
\label{KGwe}
\end{eqnarray}
In deriving Eq. (\ref{KGwe}), we have made use of the event horizon
equation (\ref{loeh}) to deal with an infinite form of $0/0$-type to
obtain a finite value
\begin{equation}
\lim_{r \rightarrow r_H}\frac{\Delta -2(r^2+a^2)\drH
+a^2\sda {\drH}^2 +{\prH}^2}{r -r_H} = 2(r_H-M) -4r_H\drH \, .
\end{equation}
By adjusting the parameter $\ka$, Eq. (\ref{KGwe}) can be reduced to
a standard wave equation near the event horizon. However, we leave
it to section 5.

\section{Dirac Equation}

To write out the explicit form of Dirac equation in the
Newman-Penrose (NP) \cite{NP} formalism, we establish the
following complex null-tetrad system that satisfies the
orthogonal conditions $\bl\cdot\bn = -\bm\cdot\overline{\bm} = 1$
\begin{eqnarray}
&&\bl=dv -a\sda d\varphi \, , ~~~\bn=\frac{\Delta}{2\Sigma}
\Big(dv -a\sda d\varphi\Big) -dr \, ,\nonumber \\
&&\bm=\frac{1}{\sqd\rho^*}\Big\{i\sta\big[adv -(r^2+a^2)d\varphi\big]
-\Sigma d\theta\Big\} \, , \nonumber\\
&&\overline{\bm}=\frac{1}{\sqd\rho}\Big\{-i\sta\Big[adv -(r^2+a^2)
d\varphi\Big] -\Sigma d\theta\Big\} \, .
\end{eqnarray}
and obtain the corresponding directional derivatives
\begin{eqnarray}
&&D=-\pr \, , ~~~\underline{\Delta}=\frac{r^2+a^2}{\Sigma}\pv
+\frac{\Delta}{2\Sigma}\pr+\frac{a}{\Sigma}\pvi \, , \nonumber\\
&&\delta=\frac{1}{\sqd\rho^*}\Big(ia\sta \pv +\pta
+\frac{i}{\sta}\pvi\Big) \, , \nonumber\\
&&\overline{\delta}=\frac{1}{\sqd\rho}\Big(-ia\sta \pv +\pta
-\frac{i}{\sta}\pvi\Big)\, .
\end{eqnarray}
It is of no difficulty to calculate the non-vanishing NP spin
coefficients of the non-stationary Kerr-Newman space-time in
the above null-tetrad as follows
\begin{eqnarray}
&&\tilde{\rho}=\frac{1}{\rho^*} \, ,
~~\epsilon=-\frac{ia\cta}{\Sigma} \, ,
~~\gamma=\frac{r\Delta}{2\Sigma^2} -\frac{r-M}{2\Sigma} \, ,
\nonumber \\
&&\tilde{\mu}=\frac{\Delta}{2\Sigma\rho^*} \, ,
~~\tau=\frac{ia\sta}{\sqd\rho^{*2}} \, ,
~~\tilde{\nu}=\frac{(\dot{M}r-Q\dot{Q})ia\sta}{\sqd\Sigma\rho} \, ,
\nonumber\\
&&\alpha=\tilde{\pi}-\beta^* \, ,
~~\tilde{\pi}=-\frac{ia\sta}{\sqd\Sigma} \, ,
~~\beta=\frac{\coa}{2\sqd\rho^*}+\frac{ira\sta}{\sqd\Sigma\rho^*} \, .
\end{eqnarray}

Inserting for the null-tetrad components of electro-magnetic potential
\begin{equation}
\bA \cdot \bn = \frac{Qr}{\Sigma} \, , ~~~~\bA \cdot \bl
= \bA \cdot \bm = -\bA \cdot \overline{\bm} = 0 \, ,
\end{equation}
and the needed spin coefficients into the four coupled Chandrasekhar-Dirac
equations \cite{CP} in the Newman-Penrose formalism
\begin{eqnarray}
&&(D+\epsilon-\tilde{\rho}+iq\bA\cdot\bl)F_1
+(\overline{\delta}+\tilde{\pi}-\alpha+iq\bA\cdot\overline{\bm})F_2
=\frac{i\mu_0}{\sqd}G_1 \, , \nonumber\\
&&(\underline{\Delta}+\mu-\gamma+iq\bA\cdot\bn)F_2
+(\delta+\beta-\tau+iq\bA\cdot\bm)F_1=\frac{i\mu_0}{\sqd}G_2 \, ,\nonumber\\
&&(D+\epsilon^*-\tilde{\rho}^*+iq\bA\cdot\bl)G_2 -(\delta+\tilde{\pi}^*
-\alpha^*+iq\bA\cdot\bm)G_1=\frac{i\mu_0}{\sqd}F_2 \, , \nonumber\\
&&(\underline{\Delta}+\mu^*-\gamma^*+iq\bA\cdot\bn )G_1
-(\overline{\delta}+\beta^*-\tau^*+iq\bA\cdot\overline{\bm})G_2
=\frac{i\mu_0}{\sqd}F_1 \, ,
\end{eqnarray}
where $\mu_0$, $q$ are the mass and charge of the Dirac particles,
respectively, we can get
\begin{eqnarray}
-\Big(\pr +\frac{r}{\Sigma}\Big)F_1
+\frac{1}{\sqd\rho}\Big({\cL} -\frac{ira\sta}{\Sigma}\Big)F_2
= \frac{i\mu_0}{\sqd}G_1 \, , \nonumber&&\\
\frac{\Delta}{2\Sigma}\Big({\cD}-\frac{ia\cta}{\Sigma}\Big)F_2
+\frac{1}{\sqd\rho^*}\Big({\cLd}-\frac{a^2\sta\cta}{\Sigma}\Big)F_1
=\frac{i\mu_0}{\sqd}G_2 \, , \nonumber&&\\
-\Big(\pr +\frac{r}{\Sigma}\Big)G_2
-\frac{1}{\sqd\rho^*}\Big({\cLd} +\frac{ira\sta}{\Sigma}\Big)G_1
= \frac{i\mu_0}{\sqd}F_2 \, ,\nonumber &&\\
\frac{\Delta}{2\Sigma}\Big({\cD}+\frac{ia\cta}{\Sigma}\Big)G_1
-\frac{1}{\sqd\rho}\Big({\cL}-\frac{a^2\sta\cta}{\Sigma}\Big)G_2
= \frac{i\mu_0}{\sqd}F_1 \, , \label{DCP} &&
\end{eqnarray}
here we have defined operators
\begin{eqnarray*}
&{\cD}&=\pr +\Delta^{-1}\Big[r-M +2iqQr +2a\pvi +2(r^2+a^2)\pv\Big] \, , \\
&{\cL}&=\pta +\frac{1}{2}\coa -\frac{i}{\sta}\pvi -ia\sta\pv \, , \\
&{\cLd}&=\pta +\frac{1}{2}\coa +\frac{i}{\sta}\pvi +ia\sta\pv \, .
\end{eqnarray*}

By substituting
$F_1=\frac{1}{\sqrt{2\Sigma}}P_1$,
$F_2=\frac{\rho}{\sqrt{\Sigma}}P_2$,
$G_1=\frac{\rho^*}{\sqrt{\Sigma}}Q_1$
and $G_2=\frac{1}{\sqrt{2\Sigma}}Q_2$
into Eq. (\ref{DCP}), we have
\begin{eqnarray}
&&-\pr P_1 +{\cL} P_2 = i\mu_0\rho^* Q_1 \, ,
~~\Delta {\cD} P_2 + {\cLd} P_1 = i\mu_0 \rho^* Q_2 \, ,
\nonumber\\
&&-\pr Q_2 -{\cLd} Q_1 = i\mu_0\rho P_2 \, ,
~~\Delta {\cD} Q_1 - {\cL} Q_2 = i\mu_0\rho P_1 \, .
\label{reDP}
\end{eqnarray}

An apparent fact is that the Chandrasekhar-Dirac equations
(\ref{reDP}) could be satisfied by setting
\begin{equation}
Q_1 \rightarrow P_2^* \, , ~~~Q_2 \rightarrow -P_1^* \, ,
~~~qQ \rightarrow -qQ \, \, .
\end{equation}

So one may deal with a pair of components $P_1$, $P_2$ only.
Eq. (\ref{reDP}) can not be decoupled except in the stationary
Kerr-Newman black hole \cite{CP} case ($M= const$) or in the
spherical symmetry Vaidya-Bonner \cite{BV} case ($a=0$). However,
to deal with the problem of Hawking radiation, one should be
concerned about the asymptotic behavior of Eq. (\ref{reDP})
near the horizon only.

First let us consider the limiting form of Eq. (\ref{reDP}) near
the event horizon. Under the transformations (\ref{trans}), Eq.
(\ref{reDP}) can be reduced to the following forms
\begin{eqnarray}
&&-\Big(\prH +ia \sta_0 \drH \Big)\spr P_1 +\Big[\Delta_H
-2(r_H^2+a^2) \drH \Big]\spr P_2 = 0 \, , \nonumber\\
&&\spr P_1 +\left(\prH -ia \sta_0 \drH \right) \spr P_2 = 0 \, ,
\label{trDPP}
\end{eqnarray}
after being taken limits $r \rightarrow r_H(v_0, \theta_0)$,
$v \rightarrow v_0$ and $\theta \rightarrow \theta_0$. It is
interesting to note that a similar form holds for $Q_1, Q_2$
also.

If the derivatives $\spr P_1$ and $\spr P_2$ in Eq. (\ref{trDPP})
are not equal to zero, the existence condition of non-trial
solutions for $P_1$ and $P_2$ is that the determinant of Eq.
(\ref{trDPP}) vanishes, which gives exactly the event horizon
equation (\ref{loeh}). The relations (\ref{trDPP}) play an
important role to eliminate the crossing-term of the first-order
derivatives in the second-order equation. It is consistent to
consider the asymptotic behavior of the first-order and
second-order Dirac equations in the meanwhile because the
four-components Dirac spinors should satisfy both of them.

Next we turn to the second-order form of Dirac equations.
A direct calculation gives
\begin{eqnarray}
&&\Big(\Delta {\cD}\pr +{\cL}{\cLd} -\mu_0^2 \Sigma\Big)P_1
= \mu_0(a\sta Q_2 -i\Delta Q_1) \nonumber\\
&&\hspace*{1cm} -ia\sta \Big[(2\dot{M}r-Q\dot{Q})\pr
+\dot{M} -2iq\dot{Q}r\Big]P_2 \, , \nonumber\\
&&\Big(\pr\Delta {\cD} +{\cLd}{\cL} -\mu_0^2 \Sigma\Big)P_2
= \mu_0(a\sta Q_1 +iQ_2) \, . \label{socd}
\end{eqnarray}

Given the GTCT in Eq. (\ref{trans}) and after some lengthy
calculations, the limiting form of Eq. (\ref{socd}), when $r$
approaches $r_H(v_0, \theta_0)$, $v$ goes to $v_0$ and $\theta$
goes to $\theta_0$, leads
\begin{eqnarray}
&&\Bigg\{\Big[\frac{r_H(1 -2\drH) -M}{\ka}+ 2\Delta_H
-2\drH (r_H^2+a^2)\Big]\spdr -2\prH \spdra  \nonumber \\
&&+2a(1 -\drH) \spdri +2(r_H^2 +a^2 -\drH a^2\sda_0)
\spdvr -(r_H -2iqQr_H \nonumber \\
&&~ -M +i a\cta_0 \drH -4r_H\drH +\prH \coa_0
+\pprH +\ddrH a^2\sda_0)\spr\Bigg\} P_1  \nonumber \\
&& =-2i(\dot{M}r_H-Q\dot{Q}) a\sta_0 \spr P_2  \nonumber\\
&& = -2i(\dot{M}r_H-Q\dot{Q}) a\sta_0 \frac{\prH
+ia \sta_0 \drH}{\Delta_H -2(r_H^2+a^2) \drH} \spr P_1 \, ,
\label{wone}
\end{eqnarray}
and
\begin{eqnarray}
&&\Bigg\{\Big[\frac{r_H(1 -2\drH) -M}{\ka} +2\Delta_H
-2\drH (r_H^2+a^2)\Big]\spdr -2\prH \spdra   \nonumber \\
&&+2a(1 -\drH) \spdri +2(r_H^2 +a^2 -\drH a^2\sda_0) \spdvr
-(M -2iqQr_H \nonumber \\
&&~ -r_H -ia\cta_0 \drH +\prH \coa_0 +\pprH
+\ddrH a^2\sda_0)\spr\Bigg\} P_2 = 0 \, ,
\label{wtwo}
\end{eqnarray}
where we have replaced the first-order derivative term
$\spr P_2$ in Eq. (\ref{wone}) by using the first
expression of relations (\ref{trDPP}).

\section{Hawking Temperature}

In order to reduce Eqs. (\ref{KGwe}), (\ref{wone}) and
(\ref{wtwo}) to a standard form of wave equation near the
event horizon, we adjust the parameter $\ka$ such
that it satisfies
\begin{equation}
\frac{r_H(1 -2\drH) -M}{\ka}+ 2\Delta_H -2\drH (r_H^2+a^2)
=r_H^2 +a^2 -\drH a^2\sda_0 \, ,
\end{equation}
which means the \lq\lq surface gravity" of the horizon is
\begin{equation}
\ka=\frac{r_H(1-2\drH)-M}{(r_H^2 +a^2 -\drH
a^2\sda_0)(1-2\drH) + 2{\prH}^2} \, , \label{temp}
\end{equation}
where we have used Eq. (\ref{loca}).

With such a parameter adjustment, these wave equations can be
recast into a combined form near the event horizon
as follows
\begin{eqnarray}
&& \Big[\spdr +2\spdvr +2\Omega_H \spdri +2C_3 \spdra \nonumber\\
&&\quad\quad\quad\quad  +2(C_2 +iC_1 +iq\Phi_H) \spr\Big] \Psi = 0 \, ,
\label{wave}
\end{eqnarray}
where
$$\Omega_H =\frac{a(1 -\drH)}{r_H^2 +a^2 -\drH a^2\sda_0} \, ,
~~~~\Phi_H =\frac{Qr_H}{r_H^2 +a^2 -\drH a^2\sda_0} \, , $$
$$ C_3 =\frac{-\prH}{r_H^2 +a^2 -\drH a^2\sda_0} \, , $$
while both $C_1$ and $C_2$ are real,
\begin{eqnarray*}
C_2&=&\frac{-1}{2(r_H^2 +a^2 -\drH a^2\sda_0)}
\Big[r_H(1-4\drH)-M +\pprH  \\
&&+\ddrH a^2\sda_0 +\prH \coa_0
+\frac{2(\dot{M}r_H-Q\dot{Q})\drH a^2\sda_0}{\Delta_H
-2\drH (r_H^2+a^2)} \Big] \, ,  \\
C_1&=&\frac{-1}{2(r_H^2 +a^2 -\drH a^2\sda_0)}
\Big[\drH a\cta_0 
-\frac{2(\dot{M}r_H-Q\dot{Q})\prH a\sta_0}{\Delta_H
-2\drH (r_H^2+a^2)}\Big] \, ,  \\
\end{eqnarray*}
for $\Psi=P_1$,
\begin{eqnarray*}
C_2&=&-\frac{M - r_H +\pprH +\ddrH a^2\sda_0
+\prH \coa_0}{2(r_H^2 +a^2 -\drH a^2\sda_0)} \, ,  \\
C_1&=&\frac{\drH a\cta_0}{2(r_H^2
+a^2 -\drH a^2\sda_0)} \, ,  \\
\end{eqnarray*}
for $\Psi=P_2$, and
\begin{eqnarray*}
C_2&=&-\frac{-2r_H\drH +\prH \coa_0 +\pprH +\ddrH a^2\sda_0
}{2(r_H^2 +a^2 -\drH a^2\sda_0)} \, ,  \\
C_1&=& 0 \, ,  \\
\end{eqnarray*}
for $\Psi=\Phi$.
We point out that the above parameter adjustment is a
crucial step to achieve a standard form of wave equation
near the event horizon, which can be viewed as an ordinary
differential equation because all coefficients in Eq.
(\ref{wave}) are regarded as finite real constants.

\section{Thermal Radiation Spectrum}

Eq. (\ref{wave}) can be treated by separating variables as
$$\Psi=R(r_*)\Theta(\theta_*)e^{i(m\varphi-\omega v_*)}$$
to the radial and angular parts
\begin{equation}
R^{\prime\prime} +2(C_0 +iC_1 +im\Omega_H +iq\Phi_H -i\omega)R^{\prime}=0 \, ,
~~~ \Theta^{\prime}=\lambda \Theta \, ,
\end{equation}
where $\lambda$ is a real constant introduced in the separation
of variables, $C_0=\lambda C_3 +C_2$. The solutions are
\begin{equation}
R=R_1e^{2i(\omega -m\Omega_H -q\Phi_H -C_1)r_* -2C_0r_*} +R_0 \, ,
~~~~ \Theta=e^{\lambda \theta_*} \, .
\end{equation}

The ingoing wave solution and the outgoing wave
solution to Eq. (\ref{wave}), are respectively,
\begin{eqnarray}
&&\Psi_{\rm in}=e^{-i\omega v_* +im\varphi +\lambda \theta_*} \, ,
\nonumber\\
&&\Psi_{\rm out}=e^{-i\omega v_* +im\varphi +\lambda \theta_*}
e^{2i(\omega -m\Omega_H -q\Phi_H - C_1)r_* -2C_0r_*}
~~~~ (r > r_H) \, .
\end{eqnarray}

The outgoing wave $\Psi_{\rm out}$ is not analytic
at the event horizon $r=r_H$, but can be analytically
continued from the outside of the hole into the inside
of the hole by the lower complex $r$-plane
$$(r -r_H) \rightarrow (r_H -r)e^{-i\pi}$$
to
\begin{equation}
\widetilde{\Psi_{\rm out}}= \Psi_{\rm out}
e^{i\pi C_0/\ka}e^{\pi(\omega -m\Omega_H -q\Phi_H - C_1)/\ka}
~~~~ (r < r_H) \, .
\end{equation}

The relative scattering probability at the event horizon is
\begin{equation}
\Big|\frac{{\Psi}_{\rm out}}{\widetilde{\Psi_{\rm out}}}\Big|^2
=e^{-2\pi(\omega -m\Omega_H -q\Phi_H - C_1)/\ka} \, .
\end{equation}
Following the method of Damour-Ruffini-Sannan's \cite{DRS},
the Hawking radiation spectra of Klein-Gordon and Dirac particles
from the black hole is easily obtained
\begin{equation}
\langle {\cal N}_{\omega} \rangle \sim
\frac{1}{e^{(\omega -m\Omega_H -q\Phi_H - C_1)/T}\pm 1} \, ,
~~~~ T=\frac{\ka}{2\pi} \, . \label{sptr}
\end{equation}
where $m$ is the azimuthal quantum number, $\Omega_H$ and $\Phi_H$
can be interpreted as the angular velocity and electro-magnetic
potential of the event horizon of the evaporating Kerr-Newman black
hole, respectively. In Eq. (\ref{sptr}), the upper plus symbol
corresponds to the Fermi-Dirac distribution, while the lower minus
symbol stands for the Bose-Einstein statistics.

\section{Spin-Rotation
Coupling Effect}

The thermal radiation spectra (\ref{sptr}) demonstrate that
the total interaction energy of particles with spin-$s$ in an
evaporating Kerr-Newman space-time is
\begin{eqnarray}
\omega_p &=& \frac{1}{r_H^2 +a^2 -\drH a^2\sda_0}
\Big[ma(1-\drH) -pa\cta_0\drH  \nonumber\\
&& +qQr_H+(s+p)\dot{M}r_H\frac{a\sta_0\prH}{\Delta_H
-2\drH(r_H^2+a^2)}\Big] \, .
\end{eqnarray}
When $p=s=0$, it corresponds to the case of scalar fields $\Psi=\Phi$;
in the case of spinor fields ($s=1/2$), $\Psi$ stands for $P_1, P_2$
when $p=1/2, -1/2$, respectively.

The energy spectrum is composed of three parts: $\omega_p = m\Omega_H
+q\Phi_H +C_1$, the first one is the rotational energy $m\Omega_H$ arising
from the coupling of the orbital angular momentum of particles with the
rotation of the black hole; the second one is the electro-magnetic
interaction energy $q\Phi_H$; another one is $C_1$ due to the coupling
of the intrinsic spin of particles and the angular momentum of the
black hole, it has no classical correspondence. From the explicit
expression of the \lq\lq spin-dependent" term $C_1$
\begin{equation}
C_1 = \frac{\Omega}{1-\drH}\Big[-p\cta_0\drH
+(s +p)\dot{M}r_H\frac{\sta_0\prH}{\Delta_H
-2\drH \big(r_H^2+a^2\big)}\Big] \, ,
\end{equation}
one can easily find that it vanishes in the case of a stationary
Kerr-Newman black hole ($M = const$, $\drH = \prH = 0$) or a
Vaidya-type black hole ($a = 0$, $\prH = 0 $, $\drH \not= 0$).

The term $C_1$ is obviously related to the helicity of particles
in different spin states, it characterizes a new effect arising
from the interaction between the spin of particles and the rotation
of an evaporating black hole. Because $\drH$ and $\prH$ describe
the evolution of the black hole in the time and the change in the
direction, we suggest that the radiative mechanism of an evaporating
Kerr-Newman black hole can be changed by the quantum rotating
ergosphere which can be viewed as a mixture of the classical
rotating ergosphere and quantum ergosphere.

\section{Conclusions}

Equations (\ref{loca}) and (\ref{temp}) give the location and
the temperature of event horizon of a non-stationary Kerr-Newman
black hole, which depend not only on the advanced time $v$ but
also on the angle $\theta$. Eq. (\ref{sptr}) shows the thermal
radiant spectra of Klein-Gordon and Dirac particles in the
non-stationary Kerr-Newman space-time. A difference between
Bosonic spectrum and Fermionic spectrum appears, that is, a
new term $C_1$ in the latter one is absent from the former one.
The new effect probably arise from the interaction between the
spin of Dirac particles and the rotation of the evaporating black
holes. The feature of this spin-rotation coupling effect is its
dependence on different helicity states of particles with spin-$1/2$
and its irrelevance to the mass of particles.

To summarize, we have dealt with Hawking radiation of Klein-Gordon
and Dirac particles in a non-stationary Kerr-Newman black hole.
The spectrum of Dirac particles displays another new effect between
the spin of the particles and the angular momentum of the hole,
which is absent from the spectrum of the Klein-Gordon particles.
This effect is due to the coupling of the intrinsic spin of particles
with the rotation of the black holes, it vanishes when the space-time
becomes a stationary Kerr black hole or a Vaidya-type spherically
symmetric black hole. This study encompasses previous ones \cite{WC2}
(when $Q = 0$) and \cite{WC3} (when $a = 0$) as special cases.

\section*{Acknowledgment}

This work is supported in part by the NSFC in China. We thank our
referee for his good advice on improving this article.


\begin{thebibliography}{99}

\bibitem{Hawk}
Hawking, S. W. (1974). {\it Nature}, {\bf 248}, 30;
(1975). {\it Commun. Math. Phys}. {\bf 43}, 199.

\bibitem{FN}
Frolov, V. P., and Novikov, I. D. (1998). {\sl Black Hole Physics:
Basic Concepts and New Developments}, (Kluwer Academic Publishers,
Dordrecht).

\bibitem{Zhao}
Zhao, Z. (1999). {\sl Thermal Properties of Black Holes and
Singularities of Space-times: Quantum Effect near the Null
Surface}, (Beijing Normal University Press, Beijing, in
Chinese).

\bibitem{HR}
Hartle, B., and Hawking, S. W. (1976). {\it Phys. Rev. D} {\bf 13}, 2188;
Wald, R. M. (1975). {\it Commun. Math. Phys}. {\bf 45}, 9;
Unruh, W. G. (1976). {\it Phys. Rev. D} {\bf 14}, 870;
Isreal, W. (1976). {\it Phys. Lett. A} {\bf 57}, 107;
Punsly, B. (1992). {\it Phys. Rev. D} {\bf 46}, 1288, 1312;
Brout, R., Massar, S., Parentani, R., and Spindel, Ph. (1995).
{\it Phys. Rep}. {\bf 260}, 329.

\bibitem{AMKP}
Khanal, U. (1983). {\it Phys. Rev. D} {\bf 28}, 1291;
Khanal, U., and Panchapakesan, N. (1981). {\it Phys. Rev. D}
{\bf 24}, 829, 835;
Ahmed, M. (1991). {\it Phys. Lett. B} {\bf 258}, 318;
Ahmed, M., and Mondal, A. K. (1995). {\it Int. J. Theor.
Phys}. {\bf 34}, 1871.

\bibitem{WC1}
Wu, S. Q., and Cai, X. (2000). {\it IL Nuovo Cimento B}
{\bf 115}, 143;
(2000). {\it Int. J. Theor. Phys}. {\bf 39}, 2215.

\bibitem{KGNK}
Zhao, Z., Dai, X. X., and Huang, W. H. (1993). {\it Acta
Astrophysica Sinica}, {\bf 13}, 299 (in Chinese);
Luo, M. W. (2000). {\it Acta Physica Sinica}, {\bf 49}, 1035 (in Chinese);
Jing, J. L., and Wang, Y. J. (1997). {\it Int. J. Theor. Phys}.
{\bf 36}, 1745.

\bibitem{Zhaoel}
Zhao, Z., Yang, C. Q., and Ren, Q. A. (1992). {\it Gen. Rel. Grav}.
{\bf 26}, 1055;
Li, Z. H., and Zhao, Z. (1993). {\it Chin. Phys. Lett}.
{\bf 10}, 126;
Zhu, J. Y., Zhang, J. H., and Zhao, Z. (1994). {\it Int. J.
Theor. Phys}. {\bf 33}, 2137;
Ma, Y., and Yang, S. Z. (1993). {\it ibid}. {\bf 32} (1993) 1237.

\bibitem{CP}
Chandrasekhar, S. (1983). {\sl The Mathematical Theory of
Black Holes}, (Oxford University Press, New York);
Page, D. (1976). {\it Phys. Rev. D} {\bf 14}, 1509.

\bibitem{WC2}
Wu, S. Q., and Cai, X. (2001). {\it Chin. Phys. Lett}.
{\bf 18}, 485;
(2001). {\it Gen. Rel. Grav}. {\bf 33}, 1181.

\bibitem{WC3}
Wu, S. Q. and Cai, X. (2001). {\it Int. J. Theor. Phys}.
{\bf 40}, 1349; (2001). {\it Mod. Phys. Lett}. A{\bf 16},
1549.

\bibitem{GHJW}
Gonzalez, C., Herrera, L., and Jimenez, J., (1979). {\it
J. Math. Phys}. {\bf 20}, 837;
Jing, J. L., and Wang, Y. J. (1996). {\it Int. J. Theor.
Phys}. {\bf 35}, 1481.

\bibitem{CKC}
Carmeli, M., and Kaye, M. (1977). {\it Ann. Phys. (NY)}
{\bf 103}, 97;
Carmeli, M. (1982). {\sl Classical Fields: General
Relativity and Gauge Theory}, (John Wiley \& Sons, New York).

\bibitem{XDY}
Xu, D. Y., (1998). {\it Class. Quant. Grav}. {\bf 15}, 153;
(1998). {\it Chin. Phys. Lett}. {\bf 15}, 706.

\bibitem{DRS}
Damour, T., and Ruffini, R. (1976). {\it Phys. Rev. D}
{\bf 14}, 332;
Sannan, S. (1988). {\it Gen. Rel. Grav}. {\bf 20}, 239.

\bibitem{NP}
Newman, E., and Penrose, R. (1962). {\it J. Math. Phys}.
{\bf 3}, 566.

\bibitem{BV}
Bonnor, W., and Vaidya, P. (1970). {\it Gen. Rel. Grav}.
{\bf 1}, 127.

\end{thebibliography}
\end{document}